\documentclass[times, doublespace]{article}
\usepackage{geometry}
\usepackage{moreverb}
\usepackage{color}
\usepackage[titletoc,toc,title]{appendix}
\usepackage[normalem]{ulem}
\usepackage{soul}
\usepackage{amsmath}
\usepackage{amssymb}

\newcommand\BibTeX{{\rmfamily B\kern-.05em \textsc{i\kern-.025em b}\kern-.08em
T\kern-.1667em\lower.7ex\hbox{E}\kern-.125emX}}

\newcommand{\R}{\mathbb{R}}
\newcommand{\p}{\mathbb{P}}

\newcommand{\bcV}{{\boldsymbol {\cal V}}}

\newcommand{\bW}{{\boldsymbol W}}
\newcommand{\bw}{{\boldsymbol w}}

\makeatletter
\newcommand*{\indep}{%
  \mathbin{%
    \mathpalette{\@indep}{}%
  }%
}
\newcommand*{\nindep}{
  \mathbin{
    \mathpalette{\@indep}{\not}    
  }
}
\newcommand*{\@indep}[2]{%
  \sbox0{$#1\perp\m@th$}
  \sbox2{$#1=$}
  \sbox4{$#1\vcenter{}$}
  \rlap{\copy0}
  \dimen@=\dimexpr\ht2-\ht4-.2pt\relax
  \kern\dimen@
  {#2}%
  \kern\dimen@
  \copy0 
} 
\makeatother

\usepackage{tikz}
\usetikzlibrary{arrows,shapes}
\tikzstyle{var}=[draw,circle,thick, text width=4mm, inner sep = 3pt]
\tikzstyle{varr}=[draw,circle,thick]
\tikzstyle{varf}=[draw=none,fill=none]
\tikzstyle{varCond}=[draw,rectangle,rounded corners=3pt, fill=gray, minimum width=2em,minimum height=2em]
\tikzstyle{edge} = [draw,thick,->,>=latex]
\tikzstyle{edge2} = [draw,thick,-]
\tikzstyle{edge3} = [draw,dashed,->,>=latex] 

\tikzset{
semi/.style={
  semicircle,
  draw,
  minimum size=2em
  }
}

\tikzstyle{varcounterfactual}=[draw, ellipse] 
\tikzstyle{varright} = [draw, semi, shape border rotate=270] 
\tikzstyle{varleft} = [draw,  semi,shape border rotate=90]

\vfuzz \hfuzz
\begin{document}

\newpage
\section*{Magnitude of selection bias in road safety epidemiology, a primer}

Marine Dufournet$^{(1)}$

\noindent{\small $^{(1)}$ Univ  Lyon, Universit\'e Claude Bernard Lyon 1, Ifsttar, UMRESTTE, UMR T\_9405, F-69373 LYON \\

\begin{abstract}
\noindent \textbf{Background:} In the field of road safety epidemiology, it is common to use responsibility analyses to assess the effect of a given factor on the risk of being responsible for an accident, among drivers involved in an accident only. Using the Structural Causal Model framework, we formally showed in previous works that the causal odds-ratio of a given factor correlated with high speed cannot be unbiasedly estimated through responsibility analyses if inclusion into the dataset depends on the crash severity. And in practice, the selection always depends on the crash severity.

\noindent \textbf{Objective:} The objective of this present work is to present numerical results to give a first quantification of the magnitude of the selection bias induced by responsibility analyses.

\noindent \textbf{Method:} We denote the following binary variables by $X$ the exposure of interest, $V$ the high speed, $F$ the driving fault, $R$ the responsibility of a severe accident, $A$ the severe accident, and $W$ a set of categorical confounders. We illustrate the potential bias by comparing the causal effect of interest of $X$ on $R$, $COR\left(X, R|W=w\right)$, and the estimable odds-ratio available in responsibility analyses, $OR\left(X, R|W=w, A=1\right)$, under a given choice of a joint distribution of $\left(X, V, F, A, W\right)$. By considering a binary exposure, and by varying a set of parameters, we describe a situation where this exposure $X$ could represent alcohol, and under additional assumptions, cannabis intoxication. 

\noindent \textbf{Results:} We confirm that the estimable odds-ratio available in responsibility analyses is a biased measure of the causal effect when $X$ is correlated with high speed $V$ and $V$ is related to the accident severity $A$. In this case, the magnitude of the bias is all the more important that these two relationships are strong. When $X$ is likely to increase the risk to drive fast $V$, the estimable odds-ratio in responsibility analyses underestimates the causal effect. When $X$ is likely to decrease the risk to drive fast $V$, the estimable odds-ratio upper estimates the causal effect. In this latter case, we especially show that a reverse direction of the estimable odds-ratio to the causal effect of interest can occur.

\noindent \textbf{Conclusion:} The values of the different causal quantities considered here are from one to five times higher (or lower) than the estimable quantity available in responsability analyses. Under additional assumptions, it is possible to observe a reverse direction. This article is the first to give a quantification of the magnitude of the bias induced by responsibility analyses and it gives new keys to well interpret the estimable odds-ratio available in such analyses. \newline

\noindent \textbf{Keywords:} causal inference ; Structural Causal Model framework; Directed Acyclic Graph; collider bias; responsibility analysis; road safety epidemiology.

\end{abstract}

\newpage
\section*{Introduction} 

Data selection mechanism poses a challenge to unbiasedly estimate the effect of a given exposure on an event. The famous ``obesity paradox" that reports an association between obesity and a decreased mortality in individuals suffering from a chronic disease is a good example of how inclusion into the dataset can create  bias \cite{Lajous_2014,viallon_re_2017}. This type of bias is known as a more general phenomenon called ``collider bias" \cite{greenland2003quantifying, rothman_modern_2008}, and it is also at play in road safety epidemiology \cite{dufournet_2017}. \\
In the field of road safety epidemiology, the collider bias occurs because the inclusion into the dataset depends in particular on the outcome of interest, the accident and even the severe accident. Indeed, road safety data are usually available when the accident occurs. Moreover, available data are often restricted to drivers and vehicles involved in severe road accidents only (\textit{e.g.}, injury or fatal accidents) \cite{amoros_estimation_2008}. To circumvent the issue of the absence of a control group, it is common to use responsibility analyses \cite{Smith_1951, Perchonok_1978, Terhune_1986, brubacher2014culpability} and to assess the effect of a given factor on the risk of being responsible for an accident, among drivers involved in an accident only. The general assumption is that non-responsible drivers represent a random sample of the general driving population that was selected to crash by circumstances beyond their control and therefore have the same risk factor profile as other drivers on the road at the same time \cite{brubacher2014culpability, Wahlberg_2007}. Then, a standard claim is that ``if this randomness assumption is met, then the risk estimate derived from a responsibility analysis would be expected to be similar to that from a case-control study" \cite{brubacher2014culpability, Wahlberg_2007}. Some authors have raised questions about the validity of these approaches \cite{Sanghavi_2013} and have sensed the presence of a residual selection bias \cite{LaumonetalBMJ} but the presence of this bias had never been formally proved yet. Consequently, in the absence of a more relevant alternative, responsibility analyses are now widely adopted in the field \cite{Asbridge_2013, Salmi_2014, Wahlberg_2009}. \\ 
Using the Structural Causal Model framework (SCM) \cite{Pearl_2000, Pearl_2010}, we formally showed in previous works \cite{dufournet_2017} that the randomness assumption does not hold in particular when inclusion in the study depends on the crash severity, through speed, and speed is affected by the considered exposure. 
In practice, the inclusion into the dataset always depends on the crash severity. On the other hand, alcohol is a well-known cause of car accidents. Since it is commonly admitted that alcohol increases the risk to drive fast, our results suggest that the estimation regarding the risk of alcohol is biased and more precisely underestimated \cite{dufournet_2017}.
\\
The objective of this article is to present numerical results to give a first accurate quantification of the magnitude of the selection bias induced by responsibility analyses. We illustrate the potential bias by comparing the causal effect of interest and the estimable odds-ratio available in responsibility analyses, under a given choice of a joint distribution of covariates involved in the mechanism leading to a severe accident. By considering a binary exposure, and by varying a set of parameters, we describe a situation where this exposure could represent alcohol, and under additional assumptions, cannabis intoxication. \\

\section{Methods}

\subsection{SCM framework and recoverability of causal effects in responsibility analyses}

Here, we give a brief description of the causal model which leads to a severe accident and remind of the issue of collider bias in responsibility analyses. Referring to Figure \ref{fig2}, our interest is in the relationship between the exposure $X$ (\textit{e.g.}, alcool or cannabis) and the binary outcome $R$ (\textit{i.e.}, responsibility for a severe accident), among drivers involved in a severe accident ($A=1$). Our causal mechanism includes $F$, the binary variable indicating whether the driver commits a driving fault, and $V$ the binary variable indicating whether a given driver drives at high speed. We consider here a simple case where $W$ denotes a unique binary confounder. We will denote by $\bcV$ the set of observable variables $\left(X, F, V, A, R, W\right)$. More details about this causal mechanism can be found in \cite{dufournet_2017}.  Moreover, because selection bias is present, we add in a specific fashion the binary variable $S$ indicating inclusion in the study \cite{Bareinboim_Pearl_2012, Bareinboim_Tian_2015}. 

\begin{figure}[!h]
\begin{center}
\begin{tikzpicture}[scale=1, auto,swap]
\node[varr] (X)at(0,0){$X$};
\node[varr] (F)at(3,0){$F$};
\node[varr] (A)at(5,0){$A$};
\node[varCond] (S)at(6,1){$S$};
\node[var] (W) at (1.5,1.5) {$W$};
\node[var] (V) at (4,1) {$V$};
\node[varr] (R) at (4,-1) {$R$};
\draw[edge] (W)--(X);
\draw[edge] (W)--(F);
\draw[thick,->,>=latex] (W) -| (A);
\draw[edge] (W)--(V);
\draw[edge] (X)--(F);
\draw[edge] (F)--(A);
\draw[edge] (F)--(R);
\draw[edge] (A)--(R);
\draw[edge] (A)--(S);
\draw[edge] (X)--(V);
\draw[edge] (V)--(F);
\draw[edge] (V)--(A);
\end{tikzpicture}
\end{center}
\caption{DAGs $G_s$ in responsibility analyses}
\label{fig2}
\end{figure}
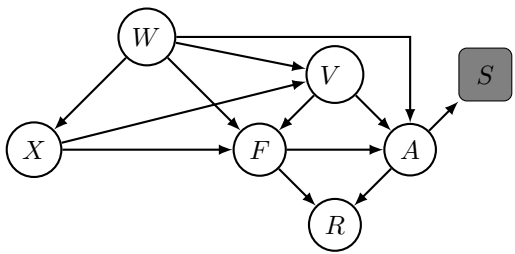

\noindent Let us precise that the binary variable $R$ is defined so that $R=F \times A$ and that we have: 

\begin{displaymath}
\left\{ \begin{array}{ll}
R=1 & \textrm{if and only if } A=1 \textrm{ and } F=1 \\
R=0 & \textrm{if } A=1 \textrm{ and } F=0 \\
R=0 & \textrm{if } A=0 \textrm{ even if } F=1. \\
\end{array} \right. \\
\end{displaymath}

\vspace{0.5cm}
\noindent In the SCM framework, the causal mechanism leading to $R$, the DAG (\textit{directed acyclic graph}) \cite{Pearl_1995,greenland_causal_1999,glymour_causal_2008}, is associated to a set of structural functions, each corresponding to one of the covariates in the DAG. This set of equations allows the definition of $R_x$, the counterfactual outcome that would have been observed in the counterfactual world where exposure would have been set to $X=x$, for $x \in \lbrace 0,1 \rbrace$. Then, causal quantities can be precisely defined. Here, we will focus on the $w$-specific causal odds-ratio of $X$ on $R$, \textit{i.e.}, the causal odds-ratio in the stratum of the population defined by $W=w$. It is defined as follows:
\begin{equation*}
COR\left(X,R|W=w\right) = \frac{\p\left(R_1=1 | W=w\right)/\p\left(R_1=0|W=w\right)}{\p\left(R_0=1 | W=w\right)/\p\left(R_0=0| W=w\right)}
\end{equation*}
Causal inference is mainly concerned with the identifiability of causal quantities. In our case, $COR\left(X, R | \bW=\bw\right)$ is identifiable if the assumptions embedded in the DAG renders it expressible in terms of the observable distribution $\p\left(\bcV=.| W=w\right)$ \cite{Bareinboim_Tian_2015}. When a selection mechanism is present, the question is the recoverability of the $w$-specific causal odds-ratio in terms of the observable distribution $\p\left({\boldsymbol {\cal V}}=.| W=w, S=1\right)$ \cite{Bareinboim_Pearl_2012, Bareinboim_Tian_2015}. If $COR\left(X, R | \bW=\bw\right)$ is recoverable, it is equal to the adjusted odds-ratio $OR\left(X,R|W=w,S=1\right)$, \textit{i.e.}, $OR\left(X,R|W=w,A=1\right)$ in our situation. If the $w$-specific causal odds-ratio is not recoverable, this also means that the estimable odds-ratio in responsibility analyses is a biased measure of causal effect. \\

\noindent In previous works, we showed that $COR\left(X, R | \bW=\bw\right)$ was not recoverable \cite{dufournet_2017}, thus $COR\left(X, R | \bW=\bw\right) \neq OR\left(X,R|W=w,A=1\right)$. Here, to assess the magnitude of bias, we compare these quantities of interest:
\begin{align*} 
COR\left(X,R|W=w\right) &= \frac{\p\left(R_1=1 | W=w\right)/\p\left(R_1=0|W=w\right)}{\p\left(R_0=1 | W=w\right)/\p\left(R_0=0| W=w\right)},  \\
OR\left(X,R|W=w,A=1\right) &= OR\left(X,F|W=w,A=1\right) \textrm{, because $F=R$ when $A=1$} \\
&= \frac{\p\left(F=1 | X=1, W=w, A=1\right)/\p\left(F=0| X=1, W=w, A=1\right)}{\p\left(F=1 | X=0, W=w, A=1\right)/\p\left(F=0| X=0, W=w, A=1\right)}
\end{align*}
We also consider: 
\begin{equation*} 
COR\left(X,F|W=w\right) = \frac{\p\left(F_1=1 | W=w\right)/\p\left(F_1=0|W=w\right)}{\p\left(F_0=1 | W=w\right)/\p\left(F_0=0| W=w\right)} \\
\end{equation*} \vspace{0.08cm}

\noindent because under additional assumptions, the causal odds-ratio $COR\left(X,R|W=w\right)$ can be well approximated by an other causal odds-ratio $COR\left(X,F|W=w\right)$. This causal odds-ratio is indeed recoverable when there is for instance no arrow pointing from $X$ to $V$ in the DAG, so that $X \indep A |\left(F,W\right)$ \cite{dufournet_2017}. Moreover, if $X \indep A |\left(F,W\right)$ holds, then $COR\left(X,R|W=w\right) \simeq COR\left(X,F|W=w\right)$ \cite{dufournet_2017}. That means that the estimable odds-ratio in responsibility analyses can approximatively and unbiasedly estimate the causal odds-ratio of interest when the considered exposure $X$ has no effect on $V$. \\

\noindent To recap, as soon as $X \nindep A |\left(F,W\right)$, the estimable odds-ratio $OR\left(X,R|W=w,A=1\right) = OR\left(X,F|W=w,A=1\right)$ is different from $COR\left(X,F|W=w\right)$ and $COR\left(X,R|W=w\right)$. Consequently, our objective is to determine the magnitude of the bias between the estimable odds-ratio and the causal odds-ratios depending on the deviation from $X \indep A |\left(F,W\right)$.

\subsection{Derivation} \label{}

Since the DAG is associated to a set of structural functions, we have to specify these functions to calculate and compare the three quantities of interest. The causal model described below is the full model consistent with Figure \ref{fig2}, \textit{i.e.}, with an arrow from $X$ to $V$. The case where there is no arrow from $X$ to $V$ will be illustrated by setting a parameter to zero in this full causal model. 
Note that we do not consider any sampling properties and hence we do not simulate data. Rather, we compare theoretical quantities under a given choice of a joint distribution of $\left(X,V,F,A,W\right)$ \cite{didelez2010assumptions}. \\

\noindent Suppose that $W$ and $X$ have distributions $f_W$ and $f_X$, respectively. While our derivations allow the variables $W$ and $X$ to take any form, the mathematics are clearer if we consider the binary case, with $\p\left(W=1\right)=p_W$ and $\p\left(X=1|W=w\right)=p_X$, and the other three variables generated by the following regression equations:
\begin{align*}
\p\left(V=1|W=w\right) = p_{V}\left(x,w\right)   &= {\rm h}\left(\alpha_0 + \alpha_X x + \alpha_W w + \alpha_{XW} xw\right) \\ 
\p\left(F=1|X=x, V=v, W=w\right) = p_{F}\left(x,v,w\right) &= {\rm h}\left(\beta_0 + \beta_X x + \beta_V v + \beta_W w + \beta_{XV} xv + \beta_{XW} xw + \beta_{VW} vw\right) \\ 
\p\left(A=1|F=f, V=v, W=w\right) = p_{A}\left(f,v,w\right) &= {\rm h}\left(\gamma_0 + \gamma_F f + \gamma_V v + \gamma_W w + \gamma_{FV} fv + \gamma_{FW} fw + \gamma_{VW} vw\right). 
\end{align*}
\noindent with ${\rm h}\left(x\right) = \left(1+ \exp\left(-x\right)\right)^{-1}$. \\
\noindent Keep in mind that $R = F \times A$, so we do not need generate the distribution of $R$ by a regression equation. \\

\noindent Consequently, the three quantities of interest $COR\left(X,R|W=w\right)$, $COR\left(X,F|W=w\right)$ and $OR\left(X,R|W=w,A=1\right)$ previously defined can be specified. Indeed, we have:

\begin{flalign*} 
&\p\left(\R_x=1 | W=w\right) = \p\left(A_x=1,F_x=1 | W=w\right) \\
&= \p\left(A_x=1,F_x=1 | X=x, W=w\right) \quad \textrm{because } \left(A_x,F_x\right) \indep X | W \\
&= \p\left(A=1,F=1 | X=x, W=w\right) \quad \textrm{by consistency} \\
&= \p\left(A=1 | F=1 , X=x, W=w\right) \p\left(F=1 | X=x, W=w\right) \\
&= \sum_{v \in \left(0,1\right)} \lbrace \p\left(A=1 | F=1, V=v, W=w\right) \p\left(F=1 | X=x, V=v, W=w\right) \p\left(V=v | X=x, W=w\right)\rbrace \\
&= p_{A}\left(1,1,w\right) p_{F}\left(x,1,w\right) p_{V}\left(x,w\right) + p_{A}\left(1,0,w\right) p_{F}\left(x,0,w\right) \left(1-p_{V}\left(x,w\right)\right)\\ 
\textcolor{white}{blabla} \\
&\p\left(F_x=1 | W=w\right) = \p\left(F_x=1 | X=x, W=w\right) \quad \textrm{because } F_x \indep X | W \\
&= \p\left(F=1 | X=x, W=w\right) \quad \textrm{by consistency} \\
&= \sum_{v \in \left(0,1\right)}\lbrace \p\left(F=1 | X=x, V=v, W=w\right) \p\left(V=v | X=x, W=w\right)\rbrace \\
&= p_{F}\left(x,1,w\right) p_{V}\left(x,w\right) + p_{F}\left(x,0,w\right) \left(1-p_{V}\left(x,w\right)\right) \\
\textcolor{white}{blabla} \\
&\p\left(R=1 | X=x, W=w, A=1\right) = \p\left(F=1 | X=x, W=w, A=1\right) \quad \textrm{because $R=F$ when $A=1$} \\
&= \frac{\p\left(F=1, X=x, W=w, A=1\right)}{\p\left(X=x, W=w, A=1\right)} \\
&= \frac{\p\left(A=1| F=1, X=x, W=w\right)\p\left(F=1, X=x, W=w\right) }{\p\left(A=1| X=x, W=w\right)\p\left(X=x| W=w\right)\p\left(W=w\right)} \\
&= \frac{\sum\limits_{v \in \left(0,1\right)} \lbrace \p\left(A=1| F=1, V=v, W=w\right) \p\left(F=1| X=x, V=v, W=w\right) \p\left(V=v| X=x, W=w\right) \rbrace}{\sum\limits_{v,f \in \left(0,1\right)}^{\textcolor{white}{n}} \lbrace \p\left(A=1| F=f, V=v, W=w\right) \p\left(F=f| X=x, V=v, W=w\right) \p\left(V=v| X=x, W=w\right) \rbrace} \\
&= [p_{A}\left(1,1,w\right) p_{F}\left(x,1,w\right) p_{V}\left(x,w\right) + p_{A}\left(1,0,w\right) p_{F}\left(x,0,w\right) \left(1-p_{V}\left(x,w\right)\right)] \\
&\quad\quad \times [p_{A}\left(1,1,w\right) p_{F}\left(x,1,w\right) p_{V}\left(x,w\right) + p_{A}\left(1,0,w\right) p_{F}\left(x,0,w\right) \left(1-p_{V}\left(x,w\right)\right) \\
&\quad\quad + p_{A}\left(0,1,w\right) \left(1-p_{F}\left(x,1,w\right)\right) p_{V}\left(x,w\right) + p_{A}\left(0,0,w\right) \left(1-p_{F}\left(x,0,w\right)\right) \left(1-p_{V}\left(x,w\right)\right)]^{-1} \\
\end{flalign*}

\noindent Throughout, we set $p_X=p_W=0.5$ and, for $\nu = 13$,
\begin{align*} 
\alpha_0 &=  -\frac{1}{2}\left(\alpha_X+ \alpha_W+\frac{1}{2}\alpha_{XW}\right)\\
\beta_0 &=  -\frac{1}{2}\left(\beta_X+ \beta_V + \beta_W + \frac{1}{2} \left(\beta_{XV} + \beta_{XW} + \beta_{VW}\right) - \nu \right) \\
\gamma_0 &=  -\frac{1}{2}\left(\gamma_F + \gamma_V + \gamma_W \frac{1}{2} \left(\beta_{FV} + \beta_{FW} + \beta_{VW}\right) - \nu \right) \\
\end{align*}
\noindent so that the prevalence $\p\left(V=1\right)=0.5$ and the prevalences of $F$ and $A$ remain inferior to $10^{-6}\%$, \textit{i.e.}, close to reality \cite{bouaoun2015road}. \newline

\noindent We remind the reader that our objective is to determine the magnitude of the bias between the estimable odds-ratio $OR\left(X,R|W=w,A=1\right)$ and the causal odds-ratios $COR\left(X,R|W=w\right)$ or $COR\left(X,F|W=w\right)$, depending on the difference from $X \indep A |\left(F,W\right)$. So, we first vary the parameters $\alpha_X$ and $\gamma_V$. For $\gamma_V$, we consider three configurations: $\gamma_V=0$, which describes the situation where there is no arrow from $V$ to $A$, $\gamma_V=1.5$ and $\gamma_V=3$. For $\alpha_X$, on one hand, we vary $\alpha_X$  from 0 to 3, which describes a situation where $X$ could represent alcohol and would increase speed. Indeed, it is commonly admitted that alcohol increases the risk to drive fast. On the other hand, we vary $\alpha_X$  from -3 to 0, which could describe a situation where $X$ would represent cannabis. The effect of cannabis on speed has not been established yet, so this result is illustrated by the situation where $\alpha_X=0$. In addition, we suppose that the cannabis could decrease speed for illustration. We also set $\beta_V = 1$ and $\gamma_F = 4$ since $V$ increases the risk of commiting a driving fault and $F$ largely increases the risk to have an accident.

\section{Results} \label{} 

\textbf{Confusion but no interaction}

\noindent In Figures \ref{fig:chap3_Xbin_alcool_acconf} and \ref{fig:chap3_Xbin_cannabis_acconf}, we can vizualise the value of $log COR\left(X,R|W=w\right)$,$log COR\left(X,F|W=w\right)$, and $log OR\left(X,R|W=w,A=1\right)$, for different values of $\alpha_X$, $\gamma_V$ and $\alpha_W$ when $\beta_W=\gamma_W=1$ and all interaction terms are set to zero. \newline

\noindent Let us first consider the case where $\alpha_W=0$ (first column of \ref{fig:chap3_Xbin_alcool_acconf} and \ref{fig:chap3_Xbin_cannabis_acconf}). The left top panel of these Figures with $\gamma_V=0$ illustrate the absence of bias between the three quantities if there had not been an arrow pointing from $V$ to $A$. In this hypothetical case, $X \indep A | \left(F,W\right)$  would hold so that $COR\left(X,F|W=w\right)$ would be recoverable and consequently $COR\left(X,R|W=w\right)\simeq OR\left(X,R|W=w, A=1\right)$. When the inclusion depends on the accident severity, \textit{i.e.} $\gamma_V \neq 0$ (second and third rows), there is some bias between $COR\left(X,R|W=w\right)$ and $OR\left(X,R|W=w,A=1\right)$ as soon as $\alpha_X \neq 0$, and the magnitude of this bias increases with $\alpha_X$ (see Figure \ref{fig:chap3_Xbin_alcool_acconf}). As expected, the higher $\gamma_V$ and $\alpha_X$ are, the bigger the magnitude of the bias is. Because we move away from  $X \indep A | \left(F,W\right)$. This last condition is also useful for the approximation of $COR\left(X,R|W=w\right)$ by $COR\left(X,F|W=w\right)$. By consequence, as soon as $\gamma_V \neq 0$, there is some bias between $COR\left(X,R|W=w\right)$ and $COR\left(X,F|W=w\right)$. However, if $\alpha_X=0$, there is no bias between the three quantities. In this case, there is no arrow pointing from $X$ to $V$ (so $V \in W$) and $X \indep A | \left(F,W\right)$ holds. Note that when $X$ increases the risk to drive fast ($\alpha_X$ varying from 0 to 3) (see Figure \ref{fig:chap3_Xbin_alcool_acconf}), the estimable odds-ratio available in responsibility analyses $OR\left(X,R|W=w,A=1\right)$ underestimates the two causal effects $COR\left(X,F|W=w\right)$ and $COR\left(X,R|W=w\right)$. Conversely, in the case where $X$ decreases the risk to drive fast ($\alpha_X$ varying from -3 to 0) (see Figure \ref{fig:chap3_Xbin_cannabis_acconf}), $OR\left(X,R|W=w,A=1\right)$ overestimates the two causal effects. Moreover, when $\alpha_X \leq -2$, a reverse direction of the estimable odds-ratio to the causal effect of interest can occur. In this latter case, the causal effect of $X$ on $R$ would be protective (log $COR\left(X,R|W=w\right) < 0$) although the estimable odds-ratio available suggests that $X$ increases the risk for being responsible of a severe accident (log $OR\left(X,R|W=w,A=1\right) > 0$). In other words, under the simple model considered here, it is possible to observe $OR\left(X,R|W=w,A=1\right) < 1$ and to conclude, if cannabis is likely to decrease the risk to drive fast, that cannabis increases the risk for being responsible for a severe accident, although it decreases this risk. \\

\begin{figure}[!h]
\caption{Causal and associational effect on the log scale in the case where $\beta_V = \beta_X = \beta_W = \gamma_W = \ 1$, $\gamma_F = 4$, without interaction, and for varying values of the other parameters $\alpha_X, \alpha_W, \gamma_V$. In each panel, along the $x$ axis,  $\alpha_X$ varies from $0$ to $3$} \label{fig:chap3_Xbin_alcool_acconf}
  \begin{center}
    \vspace{0.5cm}
    \includegraphics[scale=0.3]{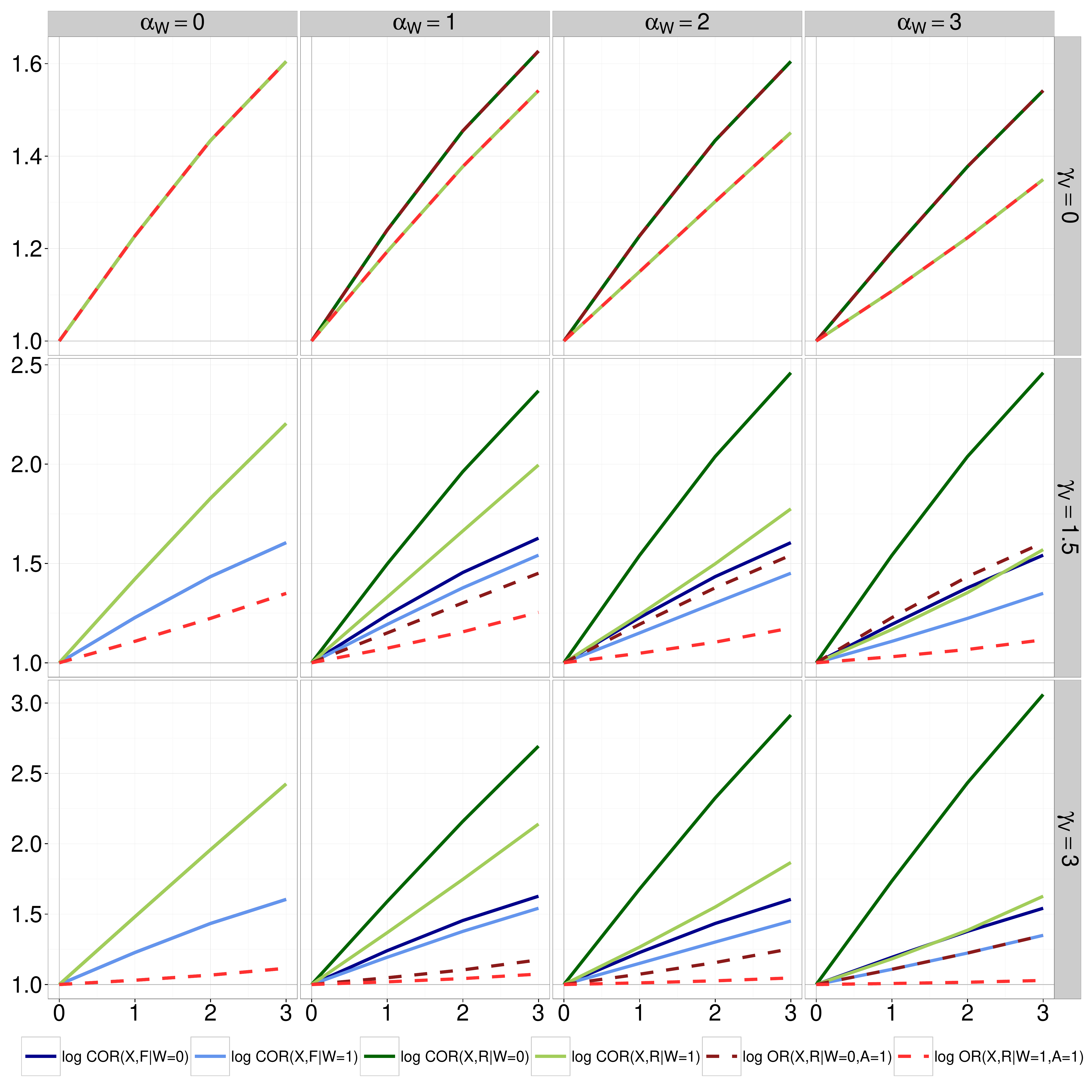} \par
    \vspace{0.5cm}
  \end{center}
\end{figure}

\begin{figure}[!h]
\caption{Causal and associational effect on the log scale in the case where $\beta_V = \beta_X = \beta_W = \gamma_W = \ 1$, $\gamma_F = 4$, without interaction, and for varying values of the other parameters $\alpha_X, \alpha_W, \gamma_V$. In each panel, along the $x$ axis,  $\alpha_X$ varies from $-3$ to $0$} \label{fig:chap3_Xbin_cannabis_acconf}
  \begin{center}
    \vspace{0.5cm}
    \includegraphics[scale=0.3]{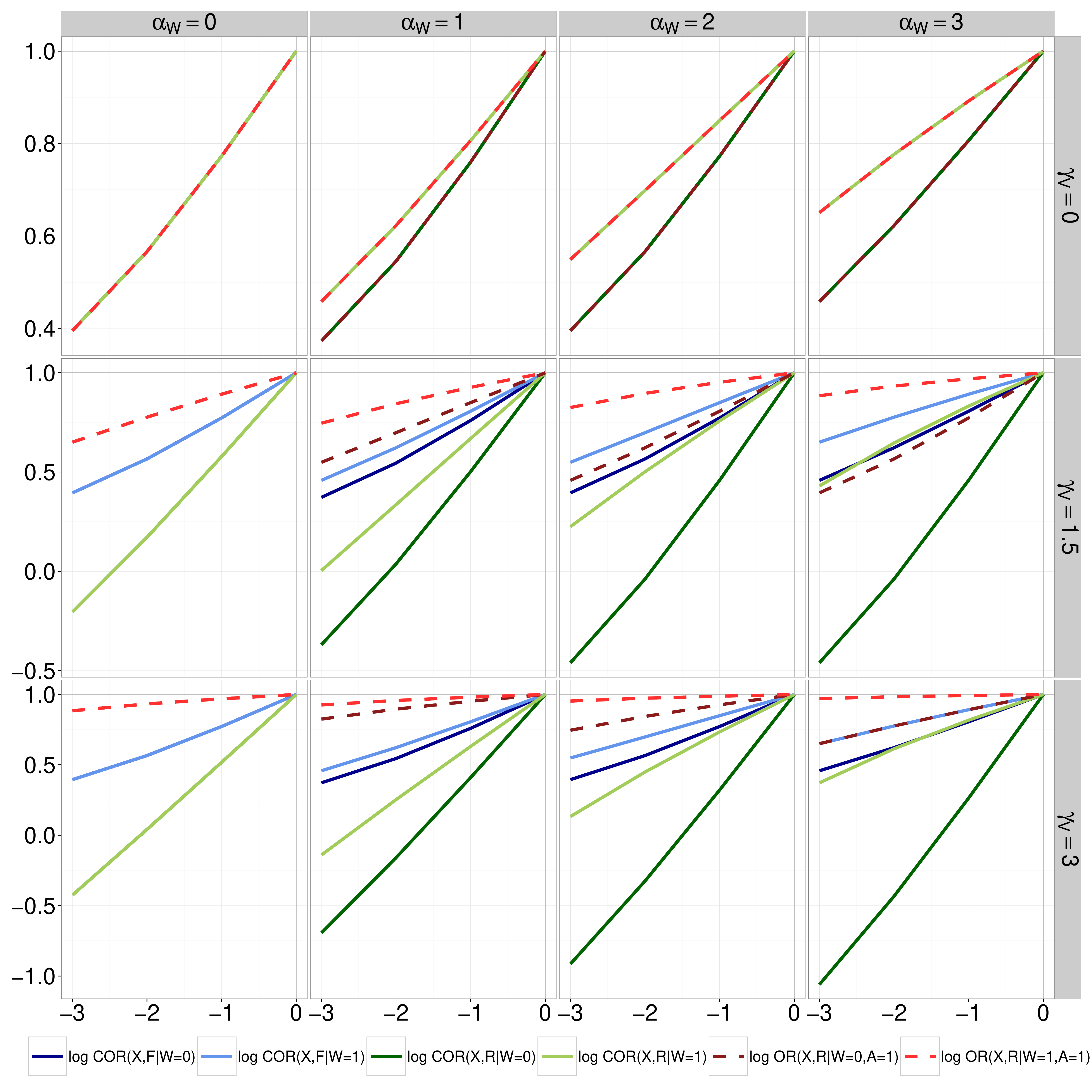} \par
    \vspace{0.5cm}
  \end{center}
\end{figure}

\noindent Now, let us consider the cases where $\alpha_W > 0$ (second, third and fourth colums of \ref{fig:chap3_Xbin_alcool_acconf} and \ref{fig:chap3_Xbin_cannabis_acconf}), which means that the binary confounder increases the risk to drive fast, as for instance to be a young driver. As expected, the bias between the three quantities increases with $\alpha_X$, whatever the values of $\gamma_V$ and $\alpha_W$, and for $w \in \left(0,1\right)$. For a given value of $\gamma_V$, the bias between $COR\left(X,R|W=0\right)$ and $COR\left(X,F|W=0\right)$ increases although it decreases between $COR\left(X,R|W=1\right)$ and $COR\left(X,F|W=1\right)$. However, the bias between $COR\left(X,F|W=w\right)$ and $OR\left(X,R|W=w,A=1\right)$ decreases whatever the stratum $W$. It seems that we move away from $X \indep A | \left(F,W\right)$ in the stratum where $\lbrace W=0 \rbrace$, \rm{e.g.} $\p\left(A=1|X=1, F=f, W=0\right)/\p\left(A=1|X=0, F=f, W=0\right)$ moves away from 1, when $\alpha_W$ increases. Conversely, we come near $X \indep A | \left(F,W\right)$ in the stratum where $\lbrace W=1 \rbrace$ \rm{e.g.} $\p\left(A=1|X=1, F=f, W=0\right)/\p\left(A=1|X=0, F=f, W=0\right)$ approaches 1. These assumptions are confirmed by computing the relative risks in each stratum of $W$ (see Appendix Table \ref{tab1}). 

\noindent Finally, as soon as $\gamma_V \neq 0$, the estimable odds-ratio available in responsibility analyses $OR\left(X,R|W=w,A=1\right)$ better estimates $COR\left(X,F|W=w\right)$ than $COR\left(X,R|W=w\right)$. For instance, when the exposure $X$ is likely to increase $V$ ($\alpha_X > 0$) (see Figure \ref{fig:chap3_Xbin_alcool_acconf}), $COR\left(X,F|W=w\right) \leq  COR\left(X,R|W=w\right)$ (see Appendix \ref{App1} for the proof). On the other hand, $OR\left(X,R|W=w,A=1\right) \leq  COR\left(X,F|W=w\right)$ and $OR\left(X,R|W=w,A=1\right) \leq  COR\left(X,R|W=w\right)$ according to the numerical illustration. It is therefore not surprising that $OR\left(X,R|W=w,A=1\right)$ better estimates $COR\left(X,F|W=w\right)$ than $COR\left(X,R|W=w\right)$. \\

\newpage
\noindent \textbf{Confusion and interactions}

\noindent In the presence of a confounder, the variation of interaction terms $\beta_{XW}, \beta_{VW}, \gamma_{FW}, \gamma_{FV}$ marginally changes the value and the magnitude of the bias. That is the reason why we do not present these results and we set these iteraction terms to 0. We then study the variation of the bias in the presence of interaction terms $\alpha_{XW}$ and $\beta_{XV}$ for $\alpha_W = 2$, $\beta_W = 1$ and $\gamma_W = 1$. \newline

\begin{figure}[!h] 
\caption{Causal and associational effect on the log scale in the case where $\beta_V = \beta_X = \beta_W = \gamma_W = \ 1$, $\gamma_F = 4, \alpha_W=2$, and for varying values of the other parameters $\alpha_X, \alpha_{XW}, \gamma_V$. In each panel, along the $x$ axis,  $\alpha_X$ varies from 0 to 3} \label{fig:chap3_Xbin_alcool_inter_alphaXU}
  \begin{center}
    \vspace{0.5cm}
    \includegraphics[scale=0.30]{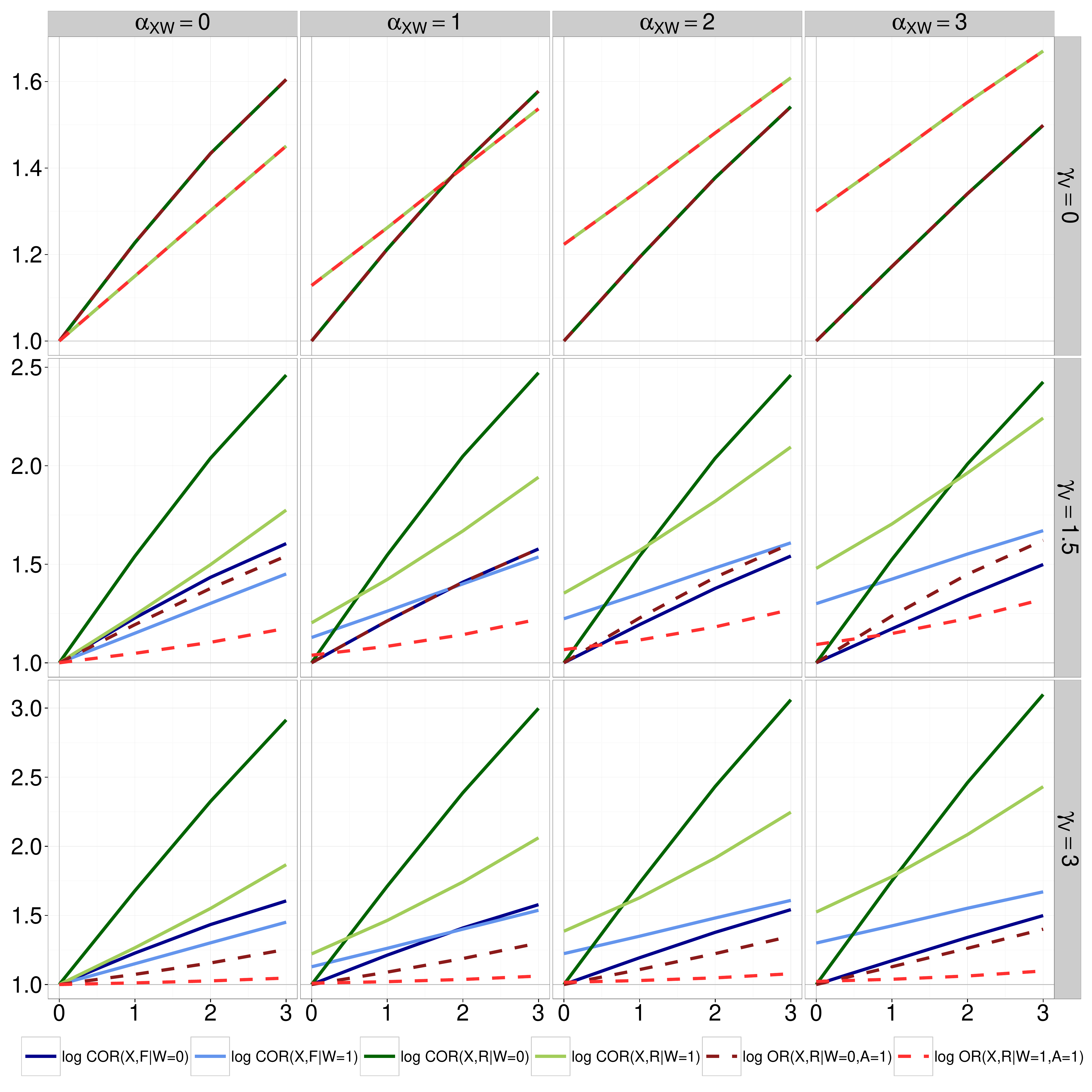} \par
    \vspace{0.5cm}
  \end{center}
\end{figure}

\noindent The interaction terms on high speed $\alpha_{XW}$ can illustrate an interaction between alcohol or cannabis consumption with the fact of being a young driver. For instance, it may be assumed that young drivers who consume alcohol take excessive risks, and drive faster than other drivers. \\
According to Figures \ref{fig:chap3_Xbin_alcool_inter_alphaXU} and \ref{fig:chap3_Xbin_cannabis_inter_alphaXU}, the variation and the magnitude of bias between the three effects depend on the stratum $W$. When $\alpha_X$ varies from 0 to 3, the bias seems to be constant in the stratum where $\lbrace W=0 \rbrace$. However, the bias increases with $\alpha_{XW}$ in the stratum where $\lbrace W=1 \rbrace$. Indeed, we deviate from $X \indep A | \left(F,W\right)$ in the stratum where $\lbrace W=1 \rbrace$ although the difference to $X \indep A | \left(F,W\right)$ is constant when $\lbrace W=0 \rbrace$ (see Appendix Table \ref{tab1}). As  expected, the variation of $\alpha_{XW}$ has a larger impact on $COR\left(X,R|W=1\right)$ and, to a lesser extent, on $COR\left(X,F|W=1\right)$, than on other effects. As already mentioned, we move away from $X \indep A | \left(F,W\right)$ when $\alpha_{XW}$ increases in the stratum where $\{W=1\}$. Consequently, $\p\left(A=1|F=f, X=1, W=1\right)/\p\left(A=1|F=f, X=0, W=1\right)$ is more and more superior than $COR\left(X,F|W=1\right)$. Moreover, note that there is a bias between the three quantities in the presence of $\alpha_{XW}$ even when $\alpha_X = 0$ in the stratum $\{W=1\}$. In this case, $X \nindep A | \left(F,W\right)$ and the relative risk $\p\left(A=1|X=1,F=f,W=1\right)/\p\left(A=1|X=0,F=f,W=1\right)$ is different from 1, because it depends on $\alpha_{XW}$, and not only on $\alpha_{X}$ anymore. The difference between the three quantities is all the more important that  $\alpha_{XW}$ is high. In the presence of the interaction term $\alpha_{XW}$, causal effects conditioned on $W=1$ are therefore superior to causal effects conditioned on $W=0$ when $\alpha_X = 0$. When $\alpha_X$ increases, the curves intersect. And the higher $\alpha_{XW}$ is, the higher the threshold $\alpha_X$ where the curves intersect is. \\

\noindent When $\alpha_X$ varies from -3 to 0 (see Figure \ref{fig:chap3_Xbin_cannabis_inter_alphaXU}), the interaction term $\alpha_{XW}$ induces a bias between the three quantities as soon as $\gamma_V$ and $\alpha_X \neq 0$. In the stratum where $\lbrace W=0 \rbrace$, the bias remains constant despite of the $\alpha_{XW}$ increase. In this stratum, the $\alpha_{XW}$ increase marginally changes the difference from $X \indep A | \left(F,W\right)$, \rm{e.g.} $\p\left(A=1|X=1, F=f, W=0\right)/\p\left(A=1|X=0, F=f, W=0\right)$ remains constant (see Appendix Table \ref{tab3}). However, in the stratum where $\{W=1\}$, the $\alpha_{XW}$ increase first reduces the bias between the three quantities. Indeed, $\p\left(A=1|X=1, F=f, W=1\right)/\p\left(A=1|X=0, F=f, W=1\right)$ approaches and reaches 1, and after moves away (see Appendix Table \ref{tab3}). This result is not surprising since $\alpha_X$ varies from -3 to 0 and $\alpha_{XW}$ varies from -1 to 2. Consequently, $\p\left(A=1|X=1, F=f, W=1\right)/\p\left(A=1|X=0, F=f, W=1\right) = 1$ when $\alpha_X + \alpha_{XW} = 0$.

\begin{figure}[!h]
\caption{Causal and associational effect on the log scale in the case where $\beta_V = \beta_X = \beta_W = \gamma_W = \ 1$, $\gamma_F = 4, \alpha_W=2$, and for varying values of the other parameters $\alpha_X, \alpha_{XW}, \gamma_V$. In each panel, along the $x$ axis,  $\alpha_X$ varies from -3 to 0} \label{fig:chap3_Xbin_cannabis_inter_alphaXU}
  \begin{center}
    \vspace{0.5cm}
    \includegraphics[scale=0.3]{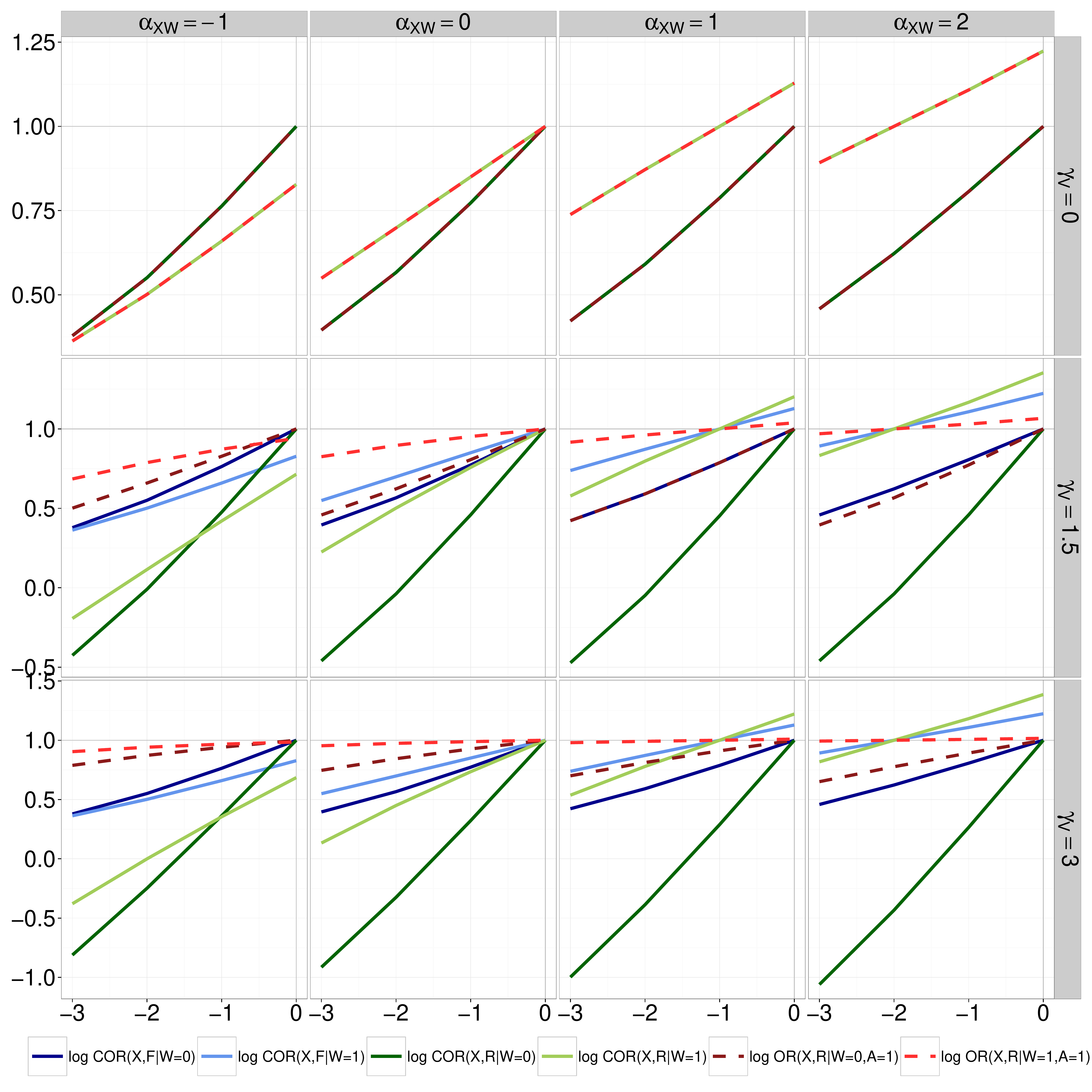} \par
    \vspace{0.5cm}
  \end{center}
\end{figure}

\noindent The interaction term $\beta_{XV}$ changes the value of the effects but not the magnitude of the bias (see Figures \ref{fig:chap3_Xbin_alcool_betaXV} and \ref{fig:chap3_Xbin_cannabis_betaXV} in Appendix \ref{App2}). Therefore, the introduction and the variation of  $\beta_{XV}$ do not change the difference from $X \indep A |\left(F,W\right)$, \rm{e.g.} the relative risk $\p\left(A=1|X=1, F=f, W=w\right)/\p\left(A=1|X=0, F=f, W=w\right)$ remains constant (see Appendix Table \ref{tab1} when $\alpha_X$ varies from 0 to 3, and \ref{tab3} when $\alpha_X$ varies from -3 to 0). Note that the introduction of interaction term $\beta_{XV}$ creates a bias between the three quantities as soon as $\gamma_V \neq 0$ and even if $\alpha_X = 0$.

\newpage
. \\
\newpage
\section{Discussion} \label{sec:ccl}

In this article, and under the causal model considered here, we show with numerical results that the estimable odds-ratio in responsibility analyses is a biased measure of the causal effect of interest as soon as selection depends on the crash severity, \rm{i.e.} $\gamma_V \neq 0$, and the exposure $X$ is correlated with high speed $V$, \rm{i.e.} $\alpha_X \neq 0$. When these two conditions are not guaranteed, the independance $X \indep A | \left(F,W\right)$ does not hold. We also show that the magnitude of the bias depends on the deviation from $X \indep A | \left(F,W\right)$ and so above all on the parameters $\alpha_X$ and $\gamma_V$. We illustrate the variation of the magnitude of the bias by varying another set of parameters. The presence of a confounder having an impact on high speed changes the deviation from $X \indep A | \left(F,W\right)$ but in a lesser extent than the variation of $\alpha_X$ or $\gamma_V$ do. Results are similar when we introduce an interaction term between $X$ and $W$ on high speed $V$. However, the introduction and the variation of the interaction between $X$ and $V$ on the fault $F$ does not change the magnitude of the bias but the value of the effects only. \\
Thanks to our numerical illustration, we are able to comment the direction and the magnitude of the bias that occurs in responsibility analyses. When $X$ is likely to increase the risk to drive fast $V$, such as alcohol, the estimable odds-ratio available in responsibility analyses underestimates the causal effect of interest, especially when $\alpha_X$ and $\gamma_V$ are high.  When $X$ is likely to decrease the risk to drive fast $V$, the estimable odds-ratio overestimates the causal effect. The relationship between the log of the effects varies from 0 to 1.5. That means that the bias between the different quantities considered here varies from 1 to 5. In the case where $X$ is likely to decrease the risk to drive fast $V$, a reverse direction of the estimable odds-ratio to the causal effect of interest can occur. In that latter case, the causal effect of $X$ on $R$ would be protective (log $COR\left(X,R|W=w\right) < 0$) although the estimable odds-ratio available suggests that $X$ increases the risk for being responsible of a severe accident (log $OR\left(X,R|W=w,A=1\right) > 0$). In other words, under the causal model considered here, it is possible to observe $OR\left(X,R|W=w,A=1\right) < 1$ and to conclude, if cannabis was likely to decrease the risk to drive fast, that cannabis would increase the risk for being responsible for a severe accident, although it would decrease that risk. Note that cannabis intoxication has no exciting effect on speed compared to alcohol but it increases the risk of commiting a driving fault $F$. \\
However, we get our results under a simple causal model with only one confounder. To be more realistic, it will be interesting to consider a set of categorical confounders for future research. Moreover, we present all results in the stratum of the population defined by $W=w$, because theoretical results on the recoverability of causal effect in the presence of selection mechanism are presented in the stratum of the population defined by $W=w$ \cite{Bareinboim_Pearl_2012, Bareinboim_Tian_2015}. Even if the recoverability of the $w$-specific $COR\left(X,R|W=w\right)$ is not sufficient to conclude on the population causal odds-ratio $COR\left(X,R\right)$, $COR\left(X,R|W=w\right)$ is still useful to derive estimates of population attributable fractions (PAF) (see Appendix C in \cite{dufournet_2017}). Consequently, our numerical results are also useful to discuss recent estimations of attributable fractions, as the one estimated by \cite{martin2017cannabis} for instance. \\
Indeed, in the recent study named ActuSAM, \cite{martin2017cannabis} compare the effect of alcohol consumption and the effect of cannabis intoxication on the risk of being responsible among drivers invoved in a fatal crash. This study concludes that drivers under the influence of alcohol are 17.8 times (12.1-26.1) more likely to be responsible for a fatal accident. Concerning cannabis intoxication, the ActuSAM study concludes that drivers under the influence of cannabis multiply their risk of being responsible for causing a fatal accident by 1.65 (1.16-2.34). By comparing the estimations (17.8 vs 1.65), and above all population attributable fractions (PAF) (27.7\% vs 4.2\%), the authors conclude that alcohol consumption remains the main problem on the french roads. Our previous results are useful to discuss the estimated attributable fractions. For instance, in the case of alcohol, it is commonly admitted that alcohol increases the risk to drive fast. Thus, our results suggest that the attributable fraction concerning alcohol would be underestimated. Regarding cannabis intoxication, the effect of cannabis on speed has not been established yet. In this situation, our results suggest that the attributable fraction estimated in the ActuSAM is unbiased, provided that the relevant confounders have been taken into account. Our new findings do not negate the global conclusion of the study, because alcohol remains a major health problem on french roads. \\
The ActuSAM study also considers different blood alcohol concentrations.  The higher the blood alcohol concentration, the higher the risk of being responsible for a fatal crash. Simple results (without confounders and interaction) regarding the magnitude of the bias when $X$ is a categorical variable are available in Appendix \ref{App3}. Results are similar to the binary case where $X$ is likely to increase the risk to drive fast $V$. Indeed, the estimable odds-ratios also underestimates the causal effects. For the highest blood alcohol concentrations, though the bias is less important because we suppose that the highest blood alcohol concentrations have a lesser impact on the risk to drive fast than the lowest concentrations. This assumption is based on the results from a linear regression model of speed on alcohol and cannabis adjusted on the same set of confounders than the ones choosen in the ActuSAM study:  age, gender, vehicle category and time of accident. Nevertheless, this assumption is questionable because it is based on biased estimations. Indeed, collider bias still occurs in the relationship between $X$ and $V$ after conditioning on $A$, if we implement the model on the case and on control groups. We have knowledge that unbiased estimations are obtained by realising the modelisation of $V$ on the control subpopulation \cite{vanderweele_odds_2010}.  However, there are not enough drunk drivers in the subpopulation of nonresponsible drivers to implement that solution. \\
Consequently, this warrants further research on the magnitude of bias. On the one hand, it would be interesting to study the magnitude of the bias when one wants to assess the effect of a given exposure on speed. We could compare the three quantities: $COR\left(X,V|W=w\right)$, $OR\left(X,V|W=w,A=1\right)$ and $OR\left(X,V|W=w,A=1\right)$. On the other hand, it would be useful to compare the magnitude of the bias between $COR\left(X,R|W=w,A=1\right)$ and $OR\left(X,R|W=w,A=1\right)$. Indeed, we have shown that $OR\left(X,R|W=w,A=1\right) \neq COR\left(X,R|W=w\right)$ but $OR\left(X,R|W=w,A=1\right)$ is also not equal to $COR\left(X,R|W=w,A=1\right)$ \cite{dufournet_2017}. Similarly to what we presented in \cite{viallon_re_2017}, we would have to derive an analytic expression of $COR\left(X,R|W=w,A=1\right)$, and this type of derivation could be also used in other fields than road safety epidemiology \cite{vansteelandt2017asking}. 

\section*{Acknowledgement} \label{sec:ack}
I would like to express my gratitude to Vivian Viallon, my research supervisor, for his guidance, methodological help and advice to build this research work. My grateful thanks are also extended to Jean-Louis Martin and Charles Fortier for their advice and proofreading. 

\newpage
\bibliographystyle{ieeetr}
\bibliography{biblio}

\begin{thebibliography}{10}

\bibitem{Lajous_2014}
M.~Lajous, A.~Bijon, G.~Fagherazzi, M.-C. Boutron-Ruault, B.~Balkau,
  F.~Clavel-Chapelon, and M.~A. Hern{\'a}n, ``Body mass index, diabetes, and
  mortality in french women: explaining away a ``paradox'','' {\em Epidemiology
  (Cambridge, Mass.)}, vol.~25, no.~1, p.~10, 2014.

\bibitem{viallon_re_2017}
V.~Viallon and M.~Dufournet, ``Re: Collider bias is only a partial explanation
  for the obesity paradox,'' {\em Epidemiology}, vol.~28, no.~5, pp.~e43--e45,
  2017.

\bibitem{greenland2003quantifying}
S.~Greenland, ``Quantifying biases in causal models: classical confounding vs
  collider-stratification bias,'' {\em Epidemiology}, vol.~14, no.~3,
  pp.~300--306, 2003.

\bibitem{rothman_modern_2008}
K.~J. Rothman, S.~Greenland, and T.~L. Lash, {\em Modern Epidemiology}.
\newblock Lippincott Williams \& Wilkins, 2008.

\bibitem{dufournet_2017}
M.~Dufournet, E.~Lanoy, J.-L. Martin, and V.~Viallon, ``Causal inference to
  formalize responsibility analyses,'' {\em Journal of causal inference (under
  revision)}, 2017.

\bibitem{amoros_estimation_2008}
E.~Amoros, J.-L. Martin, and B.~Laumon, ``Estimation de la morbidit{\'e}
  routi{\`e}re, {France}, 1996-2004,'' {\em Bulletin {\'e}pid{\'e}miologique
  hebdomadaire}, vol.~19, pp.~157--160, 2008.

\bibitem{Smith_1951}
H.~Smith and R.~Popham, ``Blood alcohol levels in relation to driving.,'' {\em
  Canadian Medical Association Journal}, vol.~65, no.~4, pp.~325--328, 1951.

\bibitem{Perchonok_1978}
K.~Perchonok, ``Identification of specific problems and countermeasures targets
  for reducing alcohol related casualties.'' Washington, DC: National Highway
  Traffic Safety Administration, U.S. Department of Transportation, 1978.

\bibitem{Terhune_1986}
K.~Terhune, ``Problems and methods in studying drug crash effects,'' {\em
  Alcohol, Drugs, and Driving}, vol.~2, no.~3-4, 1986.

\bibitem{brubacher2014culpability}
J.~Brubacher, H.~Chan, and M.~Asbridge, ``Culpability analysis is still a
  valuable technique,'' {\em International Journal of Epidemiology}, vol.~43,
  no.~1, pp.~270--272, 2014.

\bibitem{Wahlberg_2007}
A.~E. Wahlberg and L.~Dorn, ``Culpable versus non-culpable traffic accidents;
  what is wrong with this picture?,'' {\em Journal of Safety Research},
  vol.~38, no.~4, pp.~453--459, 2007.

\bibitem{Sanghavi_2013}
P.~Sanghavi, ``Commentary: Culpability analysis won't help us understand crash
  risk due to cell phones,'' {\em International Journal of Epidemiology},
  vol.~42, no.~1, pp.~267--269, 2013.

\bibitem{LaumonetalBMJ}
B.~Laumon, B.~Gadegbeku, J.-L. Martin, and M.-B. Biecheler, ``Cannabis
  intoxication and fatal road crashes in france: population based case-control
  study,'' {\em {BMJ}}, vol.~331, no.~7529, p.~1371, 2005.

\bibitem{Asbridge_2013}
M.~Asbridge, J.~R. Brubacher, and H.~Chan, ``Cell phone use and traffic crash
  risk: a culpability analysis,'' {\em International Journal of Epidemiology},
  vol.~42, no.~1, pp.~259--267, 2013.

\bibitem{Salmi_2014}
L.~R. Salmi, L.~Orriols, and E.~Lagarde, ``Comparing responsible and
  non-responsible drivers to assess determinants of road traffic collisions:
  time to standardise and revisit,'' {\em Injury Prevention}, vol.~20,
  pp.~380--386, Jan. 2014.

\bibitem{Wahlberg_2009}
A.~E. Wahlberg, ``The determination of fault in collisions,'' in {\em Driver
  Behaviour and Accident Research Methodology: Unresolved Problems.} (G.~Dorn,
  L~Matthews and I.~Glendon, eds.), pp.~101--120, Farnham, Surrey, IK: Ashgate
  Publishing, 2009.

\bibitem{Pearl_2000}
J.~Pearl, {\em Causality: models, reasoning, and inference}.
\newblock Cambridge, U.K. ; New York: Cambridge University Press, 2000.

\bibitem{Pearl_2010}
J.~Pearl, ``An introduction to causal inference,'' {\em The International
  Journal of Biostatistics}, vol.~6, no.~2, 2010.

\bibitem{Bareinboim_Pearl_2012}
E.~Bareinboim and J.~Pearl, ``Controlling selection bias in causal inference,''
  {\em Proceedings of The Fifteenth International Conference on Artificial
  Intelligence and Statistics (AISTATS 2012); JMLR}, vol.~22, pp.~100--108,
  2012.

\bibitem{Bareinboim_Tian_2015}
E.~Bareinboim and J.~Tian, ``Recovering causal effects from selection bias.,''
  in {\em Proceedings of the 29th AAAI Conference on Artificial Intelligence,
  AAAI}, pp.~3475--3481, 2015.

\bibitem{Pearl_1995}
J.~Pearl, ``Causal diagrams for empirical research,'' {\em Biometrika},
  vol.~82, pp.~669--688, Jan. 1995.

\bibitem{greenland_causal_1999}
S.~Greenland, J.~Pearl, and J.~M. Robins, ``Causal {Diagrams} for
  {Epidemiologic} {Research},'' {\em Epidemiology}, vol.~10, pp.~37--48, Jan.
  1999.

\bibitem{glymour_causal_2008}
M.~Glymour and S.~Greenland, ``Causal diagrams,'' in {\em Modern epidemiology},
  pp.~183--209, 3rd ed. lippincott williams \& wilkins~ed., 2008.

\bibitem{didelez2010assumptions}
V.~Didelez, S.~Meng, and N.~A. Sheehan, ``Assumptions of iv methods for
  observational epidemiology,'' {\em Statistical Science}, pp.~22--40, 2010.

\bibitem{bouaoun2015road}
L.~Bouaoun, M.~M. Haddak, and E.~Amoros, ``Road crash fatality rates in france:
  A comparison of road user types, taking account of travel practices,'' {\em
  Accident Analysis \& Prevention}, vol.~75, pp.~217--225, 2015.

\bibitem{martin2017cannabis}
J.-L. Martin, B.~Gadegbeku, D.~Wu, V.~Viallon, and B.~Laumon, ``Cannabis,
  alcohol and fatal road accidents,'' {\em PLoS one}, vol.~12, no.~11,
  p.~e0187320, 2017.

\bibitem{vanderweele_odds_2010}
T.~J. VanderWeele and S.~Vansteelandt, ``Odds {Ratios} for {Mediation}
  {Analysis} for a {Dichotomous} {Outcome},'' {\em American Journal of
  Epidemiology}, vol.~172, pp.~1339--1348, Dec. 2010.

\bibitem{vansteelandt2017asking}
S.~Vansteelandt, ``Asking too much of epidemiologic studies: The problem of
  collider bias and the obesity paradox,'' {\em Epidemiology}, vol.~28(5),
  no.~5, pp.~e47--e49, 2017.

\bibitem{kelly_review_2004}
E.~Kelly, S.~Darke, and J.~Ross, ``A review of drug use and driving:
  epidemiology, impairment, risk factors and risk perceptions,'' vol.~23,
  no.~3, pp.~319--344, 2004.

\bibitem{ronen_effect_2010}
A.~Ronen, H.~S. Chassidim, P.~Gershon, Y.~Parmet, A.~Rabinovich,
  R.~Bar-Hamburger, Y.~Cassuto, and D.~Shinar, ``The effect of alcohol, {THC}
  and their combination on perceived effects, willingness to drive and
  performance of driving and non-driving tasks,'' {\em Accident Analysis \&
  Prevention}, vol.~42, no.~6, pp.~1855--1865, 2010.

\end{thebibliography}

\newpage
\appendix
\section{Appendix}

\subsection{Relative risks $\p\left(A=1|X=1, F=f, W=w\right)/\p\left(A=1|X=0, F=f,W=w\right)$}

We can compute the relative risk by deriving $\p\left(A=1|X=x, F=f, W=1\right)$. Indeed, $\p\left(A=1|X=x, F=f, W=1\right)= \sum_{v \in \left(0,1\right)} \lbrace \p\left(A=1 | X=x, V=v, W=w\right) \p\left(F=1 | X=x, V=v, W=w\right) \p\left(V=v | X=x, W=w\right)=p_{A}\left(f,1,w\right) p_{V}\left(x,w\right) + p_{A}\left(f,0,w\right) \left(1-p_{V}\left(x,w\right)\right)$. Whatever $f \in \left(0,1\right)$, the relative risks marginally vary, by consequence, we present the relative risks for $f=1$.\newline

\begin{table}[!h]
\caption{Relative risk $\p\left(A=1|X=1, F=f, W=w\right)/\p\left(A=1|X=0, F=f,W=w\right)$ for $\alpha_X = 1$ and $\gamma_V=3$ and different values of $\alpha_W$, $\alpha_{XW}$, $\beta_{XV}$} 
\label{tab1}
\begin{center}
\begin{tabular}{l|c|c|c|c}
\hline                & $\alpha_W=0$ & $\alpha_W=1$ & $\alpha_W=2$ & $\alpha_W=3$ \tabularnewline \hline        
$\lbrace W=1 \rbrace$ & 1,57 & 1,42 & 1,29 & 1,19 \tabularnewline   
$\lbrace W=0 \rbrace$ & 1,57 & 1,72 & 1,83 & 1,87 \tabularnewline \hline 
                      & $\alpha_{XW}=0$ & $\alpha_{XW}=1$ & $\alpha_{XW}=2$ & $\alpha_{XW}=3$ \tabularnewline \hline        
$\lbrace W=1 \rbrace$ & 1,28 & 1,55 & 1,81 & 2,09 \tabularnewline   
$\lbrace W=0 \rbrace$ & 1,83 & 1,86 & 1,87 & 1,86 \tabularnewline   \hline 
                      & $\beta_{XV}=0$ & $\beta_{XV}=1$ & $\beta_{XV}=2$ & $\beta_{XV}=3$ \tabularnewline \hline        
$\lbrace W=1 \rbrace$ & 1,28 & 1,29 & 1,29 & 1,29 \tabularnewline   
$\lbrace W=0 \rbrace$ & 1,83 & 1,83 & 1,83 & 1,83 \tabularnewline   
\end{tabular}
\end{center} 
\end{table}

\begin{table}[!h]
\caption{Relative risks $\p\left(A=1|X=1, F=f, W=w\right)/\p\left(A=1|X=0, F=f,W=w\right)$ for $\alpha_X = -1$ and $\gamma_V=3$ and different values of $\alpha_{XW}$ and $\beta_{XV}$} 
\label{tab3}
\begin{center}
\begin{tabular}{l||c|c|c|c}
\hline                & $\alpha_{XW}=-1$ & $\alpha_{XW}=0$ & $\alpha_{XW}=1$ & $\alpha_{XW}=2$ \tabularnewline \hline        
$\lbrace W=1 \rbrace$ & 0,54 & 0,77 & 1,00 & 1,19 \tabularnewline   
$\lbrace W=0 \rbrace$ & 0,56 & 0,55 & 0,53 & 0,53 \tabularnewline   \hline
                      & $\beta_{XV}=0$ & $\beta_{XV}=1$ & $\beta_{XV}=2$ & $\beta_{XV}=3$ \tabularnewline \hline        
$\lbrace W=1 \rbrace$ & 0,77 & 0,77 & 0,78 & 0,78 \tabularnewline   
$\lbrace W=0 \rbrace$ & 0,55 & 0,54 & 0,55 & 0,55 \tabularnewline   
\end{tabular}
\end{center} 
\end{table}

\newpage
\subsection{Direction of bias}\label{App1}

Let us fisrt consider the case where $X$ represents alcohol and has a positive effect on the risk to drive fast ($\alpha_X$ varying from 0 to 3). On the one hand, we can show that as soon as $\gamma_V  \neq 0$, $COR\left(X,F|W=w\right) \leq  COR\left(X,R|W=w\right)$. Indeed, remind the proof of the approximation of $COR\left(X,R|W=w\right)$ by $COR\left(X,F|W=w\right)$ when $X \indep A | \left(F,W\right)$:

\begin{align*}
COR\left(X, R| \bW=\bw\right)&\approx CRR\left(X, R| \bW=\bw\right) \textrm{ if } \p\left(R_x=1| \bW=\bw\right) \textrm{ small} \\
& = \frac{\p\left(R_1 =1| \bW=\bw\right)}{\p\left(R_0 =1| \bW=\bw\right)}\\
&= \frac{\p\left(F_1 =1, A_1=1| \bW=\bw\right)}{\p\left(F_0 =1, A_0=1| \bW=\bw\right)}\\
&= \frac{\p\left(F_1 =1, A_1=1| X=1, \bW=\bw\right)}{\p\left(F_0 =1, A_0=1| X=0, \bW=\bw\right)}\\
&= \frac{\p\left(F =1, A=1| X=1, \bW=\bw\right)}{\p\left(F =1, A=1| X=0, \bW=\bw\right)} \textrm{ since } \left(A_x,F_x\right) \indep X | W \\
&= \frac{\p\left(A=1| F =1,X=1, \bW=\bw\right)\p\left(F =1|X=1, \bW=\bw\right)}{\p\left(A=1| F =1, X=0, \bW=\bw\right)\p\left(F =1|X=0, \bW=\bw\right)}\\
&= \frac{\p\left(A=1| F =1, \bW=\bw\right)\p\left(F =1|X=1, \bW=\bw\right)}{\p\left(A=1| F =1,\bW=\bw\right)\p\left(F =1|X=0, \bW=\bw\right)} \textrm{ since } X\indep A| \left(F, \bW\right) \\
&= \frac{\p\left(F =1|X=1, \bW=\bw\right)}{\p\left(F =1|X=0, \bW=\bw\right)}\\
&= \frac{\p\left(F_1 =1|\bW=\bw\right)}{\p\left(F_0 =1|\bW=\bw\right)} \textrm{ since } F_x \indep X | W \\
&= CRR\left(X, F| \bW=\bw\right)\\
&\approx COR\left(X, F| \bW=\bw\right) \textrm{ if } \p\left(F_x=1| \bW=\bw\right) \textrm{ small}
\end{align*}

\noindent When $\gamma_V  \neq 0$, $X \nindep A | \left(F,W\right)$, so that $\frac{\p\left( A=1| F =1,X=1, \bW=\bw\right)}{\p\left(A=1| F =1, X=0, \bW=\bw\right)} \neq \frac{\p\left(A=1| F =1, \bW=\bw\right)}{\p\left(A=1| F =1,\bW=\bw\right)}$. In the case where $X$ increases the risk to drive fast, $\frac{\p\left( A=1| F =1,X=1, \bW=\bw\right)}{\p\left(A=1| F =1, X=0, \bW=\bw\right)} \geq \frac{\p\left(A=1| F =1, \bW=\bw\right)}{\p\left(A=1| F =1,\bW=\bw\right)} $, and so $COR\left(X,F|W=w\right) \leq COR\left(X,R|W=w\right)$. On the other hand, according to the numerical illustrations, $OR\left(X,R|W=w,A=1\right) \leq  COR\left(X,F|W=w\right)$ and $OR\left(X,R|W=w,A=1\right) \leq  COR\left(X,R|W=w\right)$. It is therefore not surprising that $OR\left(X,R|W=w,A=1\right)$ better estimates $COR\left(X,F|W=w\right)$ than $COR\left(X,R|W=w\right)$. \newline

\noindent When $X$ represents cannabis intoxication and decreases the risk to drive fast ($\alpha_X$ varying from -3 to 0), analogous reasoning can be applied. On the one hand, $\frac{\p\left( A=1| F =1,X=1, \bW=\bw\right)}{\p\left(A=1| F =1, X=0, \bW=\bw\right)} \leq \frac{\p\left(A=1| F =1, \bW=\bw\right)}{\p\left(A=1| F =1,\bW=\bw\right)}$, and $COR\left(X,F|W=w\right) \geq COR\left(X,R|W=w\right)$. On the other hand, and according to the numerical illustrations, $OR\left(X,R|W=w,A=1\right) \geq  COR\left(X,F|W=w\right)$ and $OR\left(X,R|W=w,A=1\right) \geq  COR\left(X,R|W=w\right)$. $OR\left(X,R|W=w,A=1\right)$  better estimates $COR\left(X,F|W=w\right)$ than $COR\left(X,R|W=w\right)$. \\

\newpage
\subsection{Figures for varying values of $\beta_{XV}$} \label{App2}

\begin{figure}[!h]
\caption{Causal and associational effect on the log scale in the case where $\beta_V = \beta_X = \beta_W = \gamma_W = \ 1$, $\gamma_F = 4, \alpha_W=2$, and for varying values of the other parameters $\alpha_X, \beta_{XV}, \gamma_V$. In each panel, along the $x$ axis,  $\alpha_X$ varies from 0 to 3} \label{fig:chap3_Xbin_alcool_betaXV}
  \begin{center}
    \vspace{0.5cm}
    \includegraphics[scale=0.3]{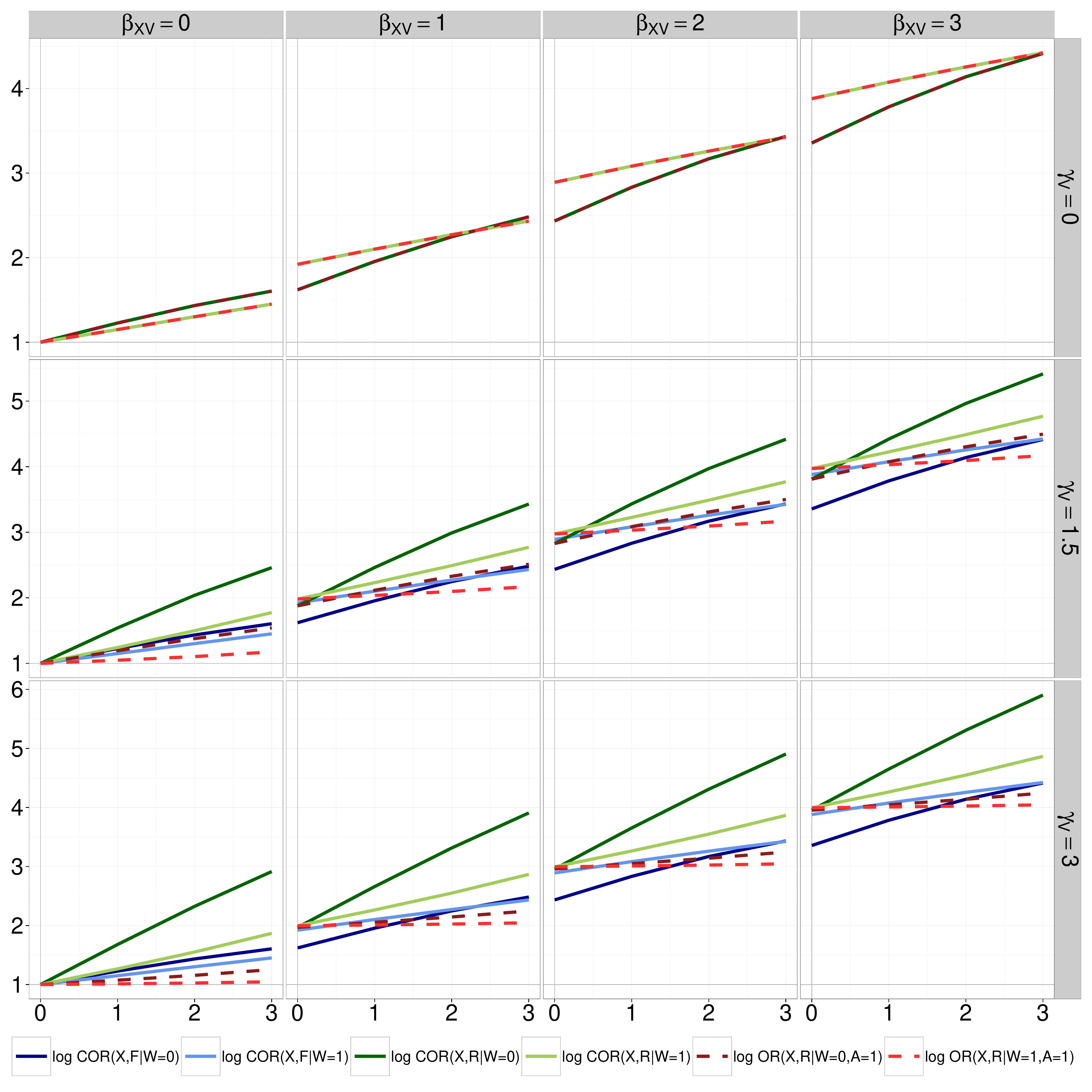} \par
    \vspace{0.5cm}
  \end{center}
\end{figure}

\newpage

\begin{figure}[!h]
\caption{Causal and associational effect on the log scale in the case where $\beta_V = \beta_X = \beta_W = \gamma_W = \ 1$, $\gamma_F = 4, \alpha_W=2$, and for varying values of the other parameters $\alpha_X, \beta_{XV}, \gamma_V$. In each panel, along the $x$ axis,  $\alpha_X$ varies from $-3$ to $0$.} \label{fig:chap3_Xbin_cannabis_betaXV}
  \begin{center}
    \vspace{0.5cm}
    \includegraphics[scale=0.3]{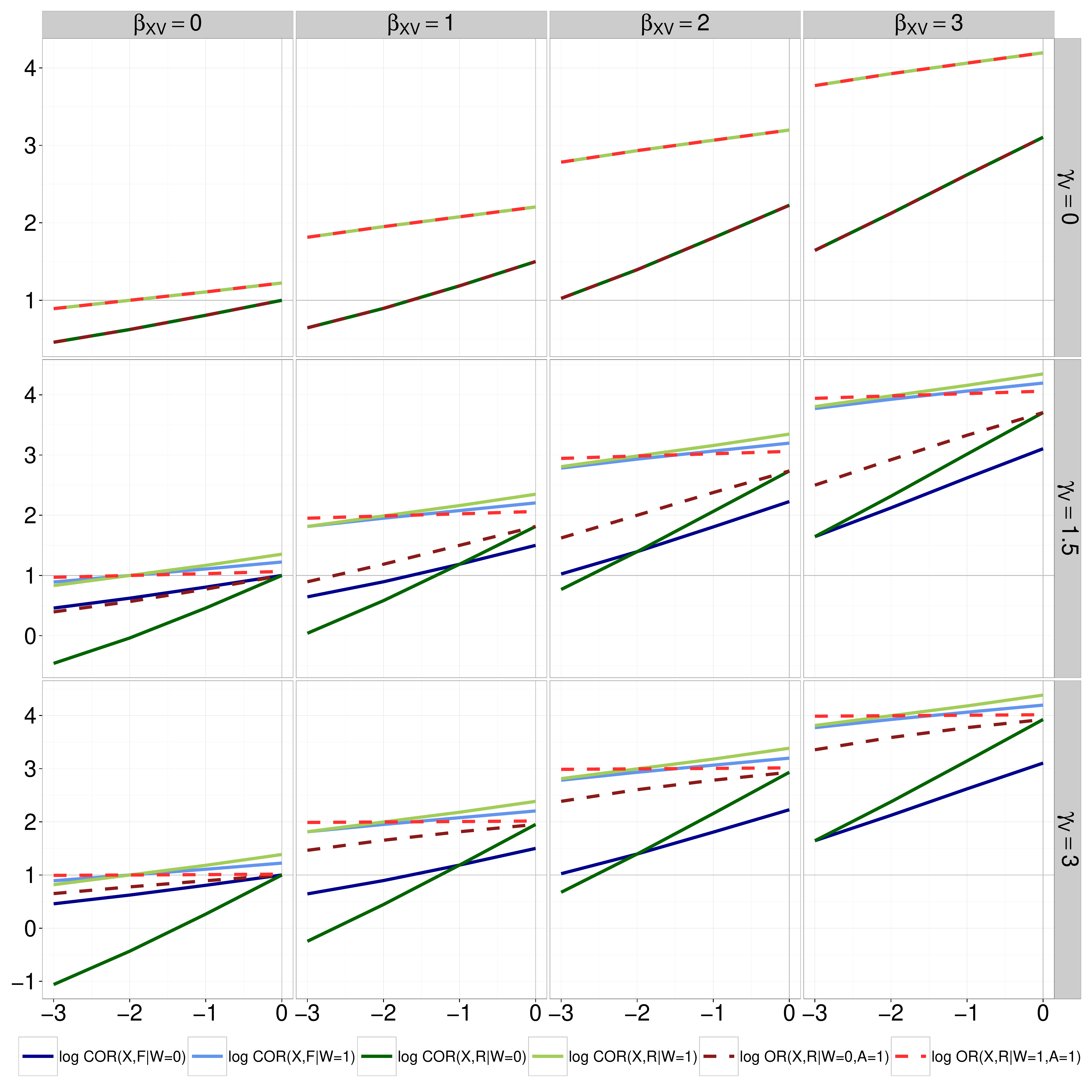} \par
    \vspace{0.5cm}
  \end{center}
\end{figure}

\newpage

\subsection{$X$ as a categorical variable} \label{App3}

\subsubsection{Derivation}

In this part, we consider the exposure $X$ as a categorical variable with $J+1$ levels $x_j \in \left(x_0,x_1,...,x_J\right)$. $X$ is transformed in $J$ binary variables $X_j$, equal to 1 if $X$ equals $x_j$, 0 otherwise for $j \geq 1$. So, for each binary variable $X_j$, we compare $COR\left(X_j,F|W=w\right)$, $COR\left(X_j,R|W=w\right)$ and $OR\left(X_j,R|W=w,A=1\right)$ defined as follow:

\begin{align*} 
COR\left(X_j,F|W=w\right) &= \frac{\p\left(F_j=1 | W=w\right)/\p\left(F_j=0|W=w\right)}{\p\left(F_0=1 | W=w\right)/\p\left(F_0=0| W=w\right)},\\
COR\left(X_j,R|W=w\right) &= \frac{\p\left(R_j=1 | W=w\right)/\p\left(R_j=0|W=w\right)}{\p\left(R_0=1 | W=w\right)/\p\left(R_0=0| W=w\right)},  \\
OR\left(X_j,R|W=w,A=1\right) &= \frac{\p\left(F=1 | X=x_j, W=w, A=1\right)/\p\left(F=0| X=x_j, W=w, A=1\right)}{\p\left(F=1 | X=x_0, W=w, A=1\right)/\p\left(F=0| X=x_0, W=w, A=1\right)}.
\end{align*}
\noindent where $F_j$ and $R_j$ are the conterfactual outcome of $F$ and $R$ respectively, that we would be observed if the exposure had been set to $X=x_j$ for $x_j \in \left(x_0, x_1, x_2, x_3, x_4\right)$.\\

\noindent As before, we note $\p\left(W=1\right)=p_W$ and $\p\left(X_j=1|W=w\right)=p_{X_j}$. The other three variables are generated by regression equations:
\begin{align*}
p_{V}\left(x_j,w\right)   &= {\rm h}\left(\alpha_0 + \alpha_{X_{j}} x_j + \alpha_W w\right) \\ 
p_{F}\left(x_j,v,w\right) &= {\rm h}\left(\beta_0 + \beta_{X_{j}} x_j + \beta_V v + \beta_W w + \beta_{VW} vw\right) \\ 
p_{A}\left(f,v,w\right) &= {\rm h}\left(\gamma_0 + \gamma_F f + \gamma_V v + \gamma_W w + \gamma_{FV} fv + \gamma_{FW} fw + \gamma_{VW} vw\right). \\
\end{align*}

\noindent Throughout, $X$ is a categorical variable with $J+1=5$ levels: $x_0$ the reference, and $x_1$, $x_2$ ,$x_3$ ,$x_4$ the other levels. So, we set $ p_{X_j} = 0.2 $ and $p_W=0.5$. For $\nu = 13$, we set 
\begin{align*} 
\alpha_0 &=  -\frac{1}{5}\left(\alpha_{X_{1}}+\alpha_{X_{2}}+\alpha_{X_{3}}+\alpha_{X_{4}} + \frac{5}{2} \alpha_W \right) \\
\beta_0 &=  -\frac{1}{5}\left(\beta_{X_{1}} + \beta_{X_{2}} + \beta_{X_{3}} + \beta_{X_{4}} + \frac{5}{2} \left(\beta_V + \beta_W\right) + \frac{5}{4} \beta_{VW} - \nu \right) \\
\gamma_0 &=  -\frac{1}{2}\left(\gamma_F + \gamma_V + \gamma_W + \frac{1}{2} \left(\beta_{FV} + \beta_{FW} + \beta_{VW}\right) - \nu \right) \\
\end{align*}
\noindent so that the prevalence $\p\left(V=1\right)=0.5$ and the prevalences of $F$ and $A$ remain inferior to $10^{-6}\%$, \textit{i.e.}, prevalences close to reality again \cite{bouaoun2015road}. \\

\noindent As in the binary case, we study the magnitude of bias between the estimable odds-ratio and the causal odds-ratios depending on the difference from $X_j \indep A |\left(F,W\right)$. So, we vary first the parameters $\alpha_{X_j}$  and $\gamma_V$ from 0 to 3. On the other hand, we set:

\begin{minipage}{0.2\linewidth}
\begin{displaymath}
\left\{ \begin{array}{lll}
\alpha_{X_{1}} \\  
\alpha_{X_{2}} = 0.80  \alpha_{X_{1}} \\
\alpha_{X_{3}} = 0.25  \alpha_{X_{1}} \\
\alpha_{X_{4}} = 0.20  \alpha_{X_{1}} \\
\end{array} \right.
\end{displaymath} 
\end{minipage}
\begin{minipage}{0.6\linewidth}
\begin{displaymath}
\left\{ \begin{array}{lll}
\beta_{X_{1}} = 1 \\  
\beta_{X_{2}} = 2.50  \beta_{X_{1}} \\
\beta_{X_{3}} = 3.50  \beta_{X_{1}} \\
\beta_{X_{4}} = 3.40  \beta_{X_{1}} 
\end{array} \right.
\end{displaymath} 
\end{minipage}

\vspace{0.5cm}
\noindent These relationships can describe a situation where $X$ would represent different blood alcohol concentrations. It is well known that alcohol increases speed for the low-dose of alcohol, and decreases it for the highest ones \cite{kelly_review_2004, ronen_effect_2010}. On the other hand, the higher blood alcohol concentration is, the higher the risk of commiting a driving fault is. Finally, we present results with $\beta_V = 1$ and $\gamma_F = 4$ always because $V$ increases the risk to commit a driving fault and $F$ largely increases the risk to have an accident.

\subsubsection{Results}

\noindent Figure \ref{fig:chap3_Xcat_txalcool_ssconf} vizualise results where $X$ is a categorical variable, with 5 levels. We find similar results as the binary case. For each level $j$, $X_j \indep A | \left(F,W\right)$ when $\gamma_V = 0$, and there is no bias between $COR\left(X_j,F|W=w\right)$, $COR\left(X_j,R|W=w\right)$ and $OR\left(X,R|W=w,A=1\right)$, whatever the value $\alpha_{X_1}$. As soon as 
$\gamma_V \neq 0$ and $\alpha_{X_1} \neq 0$, bias appear between the three quantities and the magnitude of the bias increases with $\alpha_{X_1}$, and so with $\alpha_{X_j}$. The bias are all the more important that $\alpha_{X_j}$ and $\gamma_V$ are important, because we move away from $X_j \indep A | \left(F,W\right)$. Note that the lower the relationship between $\alpha_{X_{j}}$ and $\alpha_{X_{1}}$ is, the lower the bias between the three quantities are. Finally, $OR\left(X,R|W=w,A=1\right)$  underestimates the two causal effects $COR\left(X_j,R|W=w\right)$ and $COR\left(X_j,R|W=w\right)$ since each $X_j$ variable increases the risk to drive fast. \\

\begin{figure}[!h]
\caption{Causal and associational effect for each level $j$ on the log scale in the case where  $\beta_V = \beta_{X_1}  = \ 1$, $\gamma_F = 4$ and for varying values of the other parameters $\alpha_{X_1}$ and $\gamma_V$. In each panel, along the $x$ axis,  $\alpha_{X_j}$ varies from $0$ to $3$.} \label{fig:chap3_Xcat_txalcool_ssconf}
  \begin{center}
    \vspace{0.5cm}
    \includegraphics[scale=0.3]{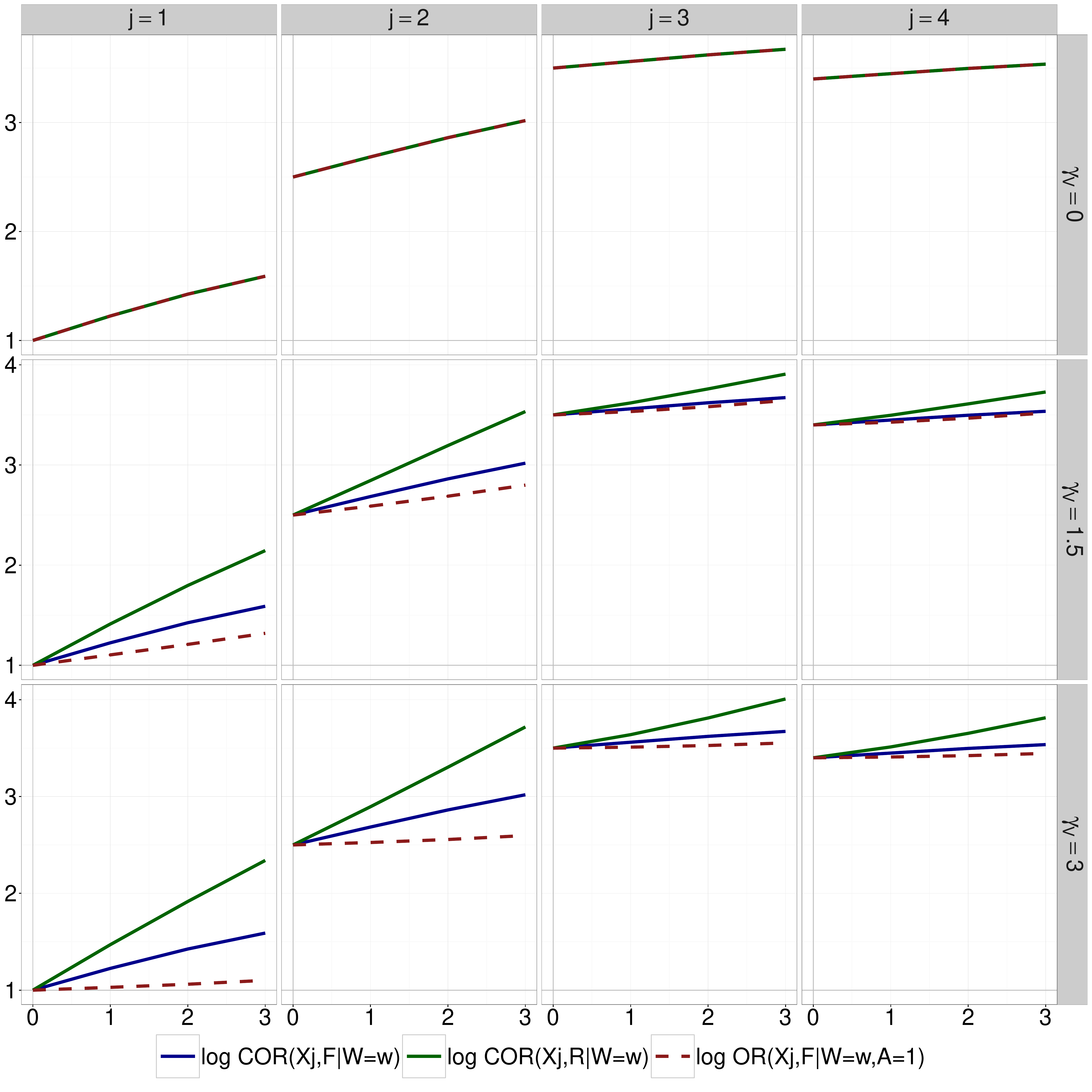} \par
    \vspace{0.5cm}
  \end{center}
\end{figure}

\end{document}